\documentclass[useAMS, usenatbib]{mn2e}
\usepackage{graphicx}
\usepackage{paralist}
\usepackage{url}
\usepackage{gensymb}
\usepackage{subfigure}
\usepackage{xcolor}
\usepackage[T1]{fontenc}

\newcommand{\argmin}{\arg\!\min}

\begin{document}
\title [Systematic biases due to calibration]
{Systematic biases in low frequency radio interferometric data due to calibration: the LOFAR EoR case} 

\author[A. H. Patil et al.]
{Ajinkya H. Patil,$^{1}$\thanks{E-mail: patil@astro.rug.nl}
Sarod Yatawatta,$^{1,2}$ Saleem Zaroubi,$^{1}$ L\'{e}on V. E. Koopmans,$^{1}$
\newauthor A. G. de Bruyn,$^{1,2}$ Vibor Jeli\'{c},$^{1,2,3}$ Benedetta Ciardi,$^{4}$ Ilian T. Iliev,$^{5}$
\newauthor  Maaijke Mevius,$^{1,2}$ Vishambhar N. Pandey,$^{1,2}$ Bharat K. Gehlot$^{1}$ \\
$^{1}$Kapteyn Astronomical Institute, University of Groningen, PO Box 800, 9700AV Groningen, the Netherlands\\
$^{2}$ASTRON, PO Box 2, NL-7990AA Dwingeloo, the Netherlands\\
$^{3}$Ru{\dj}er Bo\v{s}kovi\'{c} Institute, Bijeni\v{c}ka cesta 54, 10000 Zagreb, Croatia\\
$^{4}$Max-Planck Institute for Astrophysics, Karl-Schwarzschild-Strasse 1, D-85748 Garching bei M\"unchen, Germany\\
$^{5}$Astronomy Centre, Department of Physics \& Astronomy, Pevensey II Building, University of Sussex, Falmer, Brighton BL1 9QH, UK}

\date{}
\pagerange{}
\pubyear{}
\maketitle{}

\begin{abstract}
The redshifted 21 cm line of neutral hydrogen is a promising probe of the Epoch of Reionization (EoR). However, its detection requires a thorough understanding and control of the systematic errors. We study two systematic biases observed in the LOFAR EoR residual data after calibration and subtraction of bright discrete foreground sources. The first effect is a suppression in the diffuse foregrounds, which could potentially mean a suppression of the 21 cm signal. The second effect is an excess of noise beyond the thermal noise. The excess noise shows fluctuations on small frequency scales, and hence it can not be easily removed by foreground removal or avoidance methods. Our analysis suggests that sidelobes of residual sources due to the chromatic point spread function and ionospheric scintillation can not be the dominant causes of the excess noise. Rather, both the suppression of diffuse foregrounds and the excess noise can occur due to calibration with an incomplete sky model containing predominantly bright discrete sources. We show that calibrating only on bright sources can cause suppression of other signals and introduce an excess noise in the data. The levels of the suppression and excess noise depend on the relative flux of sources which are not included in the model with respect to the flux of modeled sources. We discuss possible solutions such as using only long baselines to calibrate the interferometric gain solutions as well as simultaneous multi-frequency calibration along with their benefits and shortcomings.
\end{abstract}

\begin{keywords}
techniques: interferometric -- methods: data analysis -- dark ages, reionization, first stars
\end{keywords}

\section{Introduction}
The first stars and galaxies formed towards the end of cosmic dark ages and  their energetic radiation is thought to have ionized matter in the Universe. The Epoch of Reionization (EoR) is the era in which matter in the intergalactic medium was transformed from being neutral to ionized. The EoR carries a wealth of information about structure formation and the first astrophysical objects in the Universe.

As hydrogen is the most abundant element in the Universe, the 21 cm transition line of neutral hydrogen is a promising probe of the EoR. The evolution of neutral hydrogen through cosmic time can be studied by observing the 21 cm line at different redshifts. The EoR is expected to have occurred between redshifts 6 and 12 \citep{Hinshaw2013, Planck2016}, which correspond to observational frequencies of 120 to 200 MHz for the redshifted 21 cm transition line. Therefore, several experiments are aiming at observing the EoR with low frequency radio telescopes including  Giant Meterwave Radio Telescope (GMRT; \citealt{Paciga2013}), Low Frequency Array (LOFAR; \citealt{vanHaarlem2013}), Murchison Widefield Array (MWA; \citealt{Tingay2013}, \citealt{Bowman2013}, \citealt{Dillon2015}, \citealt{Trott2016a}), the Donald C. Backer Precision Array for Probing the Epoch of Reionization (PAPER; \citealt{Parsons2010}, \citealt{Ali2015}), the Hydrogen Epoch of Reionization Array (HERA; \citealt{DeBoer2015}), the Square Kilometer Array (SKA; \citealt{Mellema2013}, \citealt{Koopmans2015}). 

The contamination due to the Galactic and extragalactic foreground emission is one of the primary challenges in detecting the  cosmic redshifted 21 cm emission from neutral hydrogen (hereafter referred as the 21 cm signal). The astrophysical foregrounds are either discrete sources such as radio galaxies and clusters or diffuse synchrotron and free-free emissions from our galaxy \citep{Shaver1999, DiMatteo2002, Oh2003, DiMatteo2004, Cooray2004}. These foregrounds are several orders of magnitude brighter than the expected 21 cm signal. Therefore, an accurate removal of the foregrounds while avoiding possible systematic errors is crucial for the success of EoR experiments. In this paper, we present some systematic biases observed in the residual LOFAR EoR data after calibration and subtraction of bright discrete foreground sources, investigate their origins and discuss possible solutions. 

Two important systematic biases observed in the LOFAR-EoR data after calibration and foreground subtraction are: i) a suppression of diffuse, polarized foregrounds and ii) an excess of noise. Diffuse foregrounds appear both in total and polarized intensity \citep{Jelic2014, Jelic2015}, but they are difficult to detect in total intensity (Stokes I) in presence of numerous bright discrete sources. Diffuse foregrounds are dominant in polarized intensity, because only few discrete foreground sources show polarized emission. We observe a suppression in the polarized diffuse foregrounds while subtracting discrete foreground sources. Diffuse foregrounds appear predominantly on large angular scales, which are also the most promising scales for a detection of the 21 cm signal \citep{Zaroubi2012, Chapman2013, Patil2014a}. Although one aims to detect the 21 cm signal in total intensity, a suppression of the diffuse polarized foregrounds could suggest a suppression of the 21 cm signal as well. The second systematic effect is an excess of noise beyond the thermal noise. The excess noise not only reduces sensitivity, but also causes an obstacle in the foreground removal. Several foreground removal or avoidance algorithms separate the foregrounds based on their spectral smoothness (see \cite{Chapman2015} for a review of foreground removal methods). The excess noise introduces additional random variations along frequency in the data, and hence it makes removal of foregrounds inefficient. We investigate three potential sources of the excess noise: the chromatic nature of the point spread function (PSF), ionospheric scintillation and calibration artefacts.

The response of a radio interferometer needs to be calibrated in order to correct for variations in electronics and the ionosphere. A bright compact source with known flux is needed to calibrate the gains of interferometric elements. However, few such calibrator sources are known at low radio frequencies, and it is possible that none of them might be located within the field view of an observation. One can instead use self-calibration in such cases. In self-calibration, a model of bright sources in the sky is constructed, and it is used to calibrate the gains of interferometric elements \citep{Schwab1980, Cornwell1981}. The sky model and the gain solutions are improved in an iterative manner. The traditional self-calibration obtains one gain solution for each interferometric element. However, this may not be sufficient for the new generation of telescopes with wide fields of views, where the gain might change as a function of direction. Direction dependent self-calibration is then used where the gain solutions in multiple directions are obtained \citep{vanderTol2007, Wijnholds2009}. Some EoR projects use direction dependent self-calibration for the calibration and subtraction of bright sources \citep{Mitchell2008, Yatawatta2013}. Nearby sources can be clustered together to get one solution in the respective direction \citep{Kazemi2013c}. 

The sky model in self-calibration is often imperfect due to errors in flux, position or morphology of the modelled sources \citep{Datta2009, Datta2010}. The sky model is also incomplete, because it contains only bright discrete sources and excludes faint discrete sources and diffuse emissions. Some artefacts of calibration with an incomplete sky model have been well known. These include generation of spurious source components and suppression of real components \citep{Wilkinson1988}. \cite{Grobler2014} and \cite{Wijnholds2016} considered a simple case of one bright and one faint source and provided an analytical description of how spurious sources can be generated when the faint source is excluded in the model for calibration. However, real data is more complex with many discrete sources and diffuse foregrounds. Therefore, in this paper, we rely on simulations to study effects of model incompleteness. In a similar study, \cite{Barry2016} found that excluding faint discrete sources in a sky model leads to contamination of foreground-free power spectrum modes. In this paper, we also consider effects of diffuse foregrounds, show the contamination in the observed data and discuss some solutions with their advantages and shortcomings. 

An alternative to self-calibration is redundancy calibration which does not require a priory model of the sky \citep{Noordam1982, Wieringa1992}. Therefore, the discussion in this paper does not apply to redundancy calibration. However, redundant arrays use a hybrid approach consisting of redundancy calibration followed by a sky model based calibration to resolve degeneracies of the former \citep{Ali2015, Zheng2014}.

The paper is organized as follows: in Section 2, we briefly describe the data analysis pipeline for the LOFAR-EoR project. In Section 3, we discuss systematic biases observed in the calibrated data, namely, an excess noise and suppression of the diffuse foregrounds. Detailed properties of the excess noise and possible sources of its origin are discussed in Section 4. In Section 5, we show with the help of simulations that the above two systematic biases could be artefacts of calibration with an incomplete sky model. We discuss some possible solutions to the systematic biases in Section 6, before concluding in Section 7.

\section{Observations and data processing}
\begin{figure} 
\centering
\includegraphics{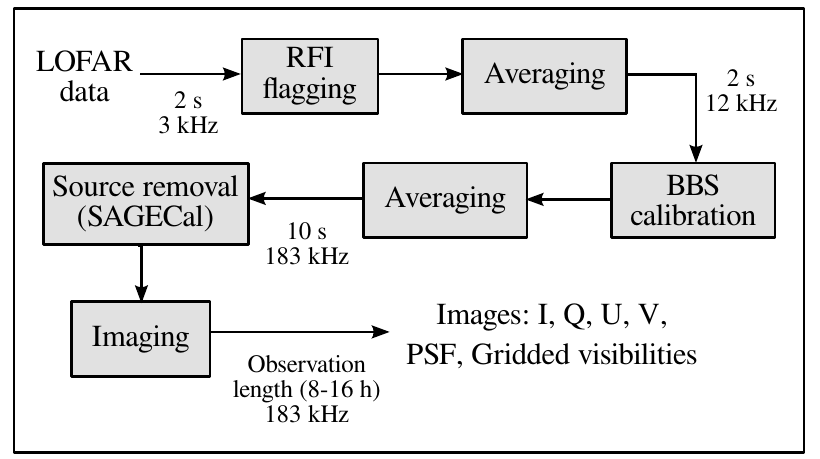} 
\caption{Block diagram of the data reduction pipeline. Time and frequency resolutions at different stages are noted. DI and DD refer to direction independent and direction dependent, respectively. The final output is a set of images of Stokes parameters (I, Q, U, V), the point spread function (PSF) and gridded visibilities.}
\label{fig:block}
\end{figure}

The data used in this paper was observed with LOFAR during observing cycle 0 (Feb-Nov 2014) and cycle 1 (Nov 2013 to May 2014). We concentrate on the primary target field of the LOFAR-EoR experiment centered on the North Celestial Pole (NCP). The NCP field was observed with 55 LOFAR High Band Antenna stations in the Netherlands, providing baselines from 68 m to 121 km, and operating in the frequency range 115-189 MHz. However, we use the data only up to 177 MHz in this paper, because the 177-189 MHz part of the bandwidth is corrupted by radio frequency interference (RFI). The frequency range 115-177 MHz corresponds to redshifts 7 to 11.35 for the 21 cm line of neutral hydrogen. Visibilities, i.e. correlations of voltages from pairs of antennas, were recorded with 2 second time resolution. The total bandwidth was divided into 195 kHz sub-bands. Each sub-band consisted of 64 channels, thereby providing a frequency resolution of 3 kHz. We observed only during night time to avoid contamination due to the solar emission and minimize ionospheric phase errors. The duration of an observation varied between 6 and 16 hours depending on the season at the time of observation. The observational details are summarized in Table~\ref{table:data}. For more information about LOFAR capabilities, the reader is referred to \cite{vanHaarlem2013}. Different steps in the processing of the observed data are summarized in the following subsections. Please see Fig. \ref{fig:block} for a block diagram of the data reduction pipeline.

\begin{table}
\centering
\caption{Observational details of the data used in this paper}
\begin{tabular}{l l}
\hline
Telescope &	LOFAR High band antennas\\
Observational period: & \\
\qquad LOFAR cycle 0 & Feb-Nov 2013\\
\qquad LOFAR cycle 1 & Nov 2013 - May 2014\\
Duration of an observation & 6-16 hours (depending on season)\\
Frequency range & 115-174 MHz\\
Field of view at 150 MHz& 3.8\degree (full width at half maximum)\\
Polarization & Linear X-Y\\
Longest baseline: & \\
\qquad LOFAR core & 3.5 km\\
\qquad LOFAR Dutch array & 121 km\\
Collecting area towards zenith: & \\
\qquad LOFAR core & 512 m${^2}$ x 48 stations\\
\qquad Dutch remote stations & 1024 m${^2}$ x 16 stations\\
Time, frequency resolution: & \\
\qquad Raw data & 2 s, 3 kHz\\
\qquad After RFI flagging & 2 s, 12 kHz\\
\qquad After calibration & 10 s, 183 kHz\\
\hline
\end{tabular}
\label{table:data}
\end{table}

\subsection{Pre-processing}
The first step in our data processing is to discard that part of the data which is affected by RFI. The RFI mitigation is performed by the software \texttt{aoflagger} \citep{Offringa2010, Offringa2012} at the highest time and frequency resolution available to minimize information loss. Two frequency channels on either edge of every sub-band are discarded to avoid edge-effects of the polyphase filter. This reduces the bandwidth of each frequency sub-band to 183 kHz. The remaining data is then averaged to 12 kHz, 2 second resolution to reduce its volume for further processing.

\subsection{Direction Independent Calibration}
Usually, a bright source with known flux can be used to calibrate the gain of each interferometric element. However, the region within the field of view at the NCP  contains not one dominant source but rather many sources with comparable fluxes e.g. NVSS 7011732+89284 with 7.2 Jy\footnote{The flux of NVSS 7011732+89284 was earlier thought to be 5.3 Jy \citep{Yatawatta2013} and was used to set the absolute flux scale. It was assumed that the source has a constant spectrum from 100 to 300 MHz. However, recent observations with LOFAR have revealed that the spectrum of the sources rises and falls within this frequency range with the correct flux of 7.2 Jy at 150 MHz (de Bruyn et al., in preparation).}, 3C61.1 with 1 to 11 Jy depending on frequency and several sources with 1 Jy apparent flux at 150 MHz. Therefore, we use 300 sources spread over the area of 10 x 10 square degrees to calibrate the average station gains over the field of view in the direction of the NCP. We use the Black Board Selfcal package \citep{Pandey2009} to obtain and apply the calibrated gain solutions for every 10 second time interval and 183 kHz bandwidth. Each station gain is described by a 2 x 2 Jones matrix for two orthogonal linear polarizations.

\subsection{Source subtraction}
Supernova remnants and radio galaxies and clusters are the discrete foreground sources observed at low radio frequencies. The brightest sources in the NCP field are about six orders of magnitude brighter than the expected 21 cm signal. Therefore, we need to remove the foreground sources with a very high accuracy to reach the required sensitivity for a signal detection. Foreground sources can be subtracted by self-calibration. However, station gains obtained towards the center of the field or the average gains over the field of view are not good enough for the entire field of view of LOFAR. Varying primary beam shapes and ionospheric effects cause direction-dependent effects \citep{Lonsdale2005, Koopmans2010, Vedantham2015}, which require obtaining gains towards multiple directions in which sources are to be removed. This is called a direction dependent calibration. We use \texttt{SAGECal} \citep{Yatawatta2009, Kazemi2011, Kazemi2013a, Kazemi2013b} to calibrate the station gains in multiple directions and ultimately subtract sources. \texttt{SAGECal} takes a sky model containing positions, fluxes and morphologies of a set of known sources as an input. It solves for the station gains in the direction of these sources by minimizing the difference between the observed data and predicted visibilities for the sky model multiplied with the estimated station gains (please see the appendix for a mathematical description of the calibration). Finally, the sources are removed by subtracting their predicted visibilities multiplied with the obtained gain solutions. It is important to note that the station gain solutions are not applied to the residual data but are only used to subtract the modelled sources. The sky model is regularly updated as we reach better sensitivities by subtracting sources and observing more data. We refer the reader to \cite{Yatawatta2013} for more details about the calibration and source subtraction in the LOFAR-EoR NCP field.

\subsection{Imaging}
Residual visibilities obtained after source subtraction are imaged using the software package \texttt{ExCon} \citep{Yatawatta2014}. We attempt to maintain the spectral smoothness of foregrounds by using uniform weighting and only the densely sampled part of the uv plane, i.e. baselines between 30 and 800 wavelengths \citep{Patil2014a}. Separate images are made for each 183 kHz wide sub-band. 

\section{Systematic biases in the data}
As a first step towards the detection of the 21 cm signal, we would like to measure the variance \citep{Patil2014a, Patil2014b} and the power spectrum \citep{Harker2010, Chapman2013} of the differential brightness temperature of the 21 cm emission as a function of redshift. Simulations in \cite{Patil2014a} show that the 21 cm signal variance can be detected with a 4$\sigma$ significance in 600 hours if all systematic errors can be controlled. However, we identify two systematic biases in the residual data after calibration and subtraction of bright discrete foreground sources, namely, an excess of noise and a suppression in diffuse foregrounds. These two problems are described in the following subsections.

\subsection{The excess noise}
An accurate determination of the statistical properties of the thermal noise such as its standard deviation and power spectrum is important. The expected standard deviation ($\sigma$) of the thermal noise in a visibility can be calculated from the system equivalent flux density (SEFD) as
\begin{equation} \label{eq:sefd}
\sigma = \frac{\mathrm{SEFD}}{\sqrt{2 \Delta \nu \Delta t}},
\end{equation}
where $\Delta \nu$ and $\Delta t$ are integration frequency bandwidth and time, respectively. The SEFD depends on the elevation of an observation. The expected SEFD of the LOFAR HBA towrads the NCP is about 4100 Jy, as derived from the empirical SEFD towards the zenith (de Bruyn et al., in prep.). For 10 second and 180 kHz integration, the noise per visibility should be 2.16 Jy. About $7 \times 10^6$ visibilities are observed over 12 hours of observation. Therefore, the thermal noise in an image made with such an observation should be about 580 $\mu$Jy. In reality, the noise in an image depends on several factors such as the fraction of the data flagged due to RFI, weights given to different visibilities during imaging, the Galactic background in the direction of observation,  calibration artefacts. A more detailed discussion about noise properties will follow in de Bruyn et al. (in prep.).

The actual thermal noise in an observation can be determined using the circular polarization data, i.e. Stokes V parameter. Most radio sources in the sky do not show circular polarization. Therefore, the Stokes V images are expected to be thermal noise dominated. There can be a small leakage of the total intensity, i.e. Stokes I, into Stokes V. Such leakage occurs because of the different projections of the two orthogonal dipoles towards the same direction in the sky. However, 
the polarization leakage for modeled sources is removed during the calibration and source removal. Furthermore, \cite{Khan2015} have shown that the Stoke I to Stokes V leakage is less than 0.003 percent. Therefore, the Stokes V images provide good estimates of the noise properties. The root mean square (RMS) of the Stokes V noise in our data is about 0.9 mJy for a 13 hours and 195 kHz (one sub-band) integration at 150 MHz in uniform-weighted images of 3 arcmin resolution.

\begin{figure}
\centering
\includegraphics{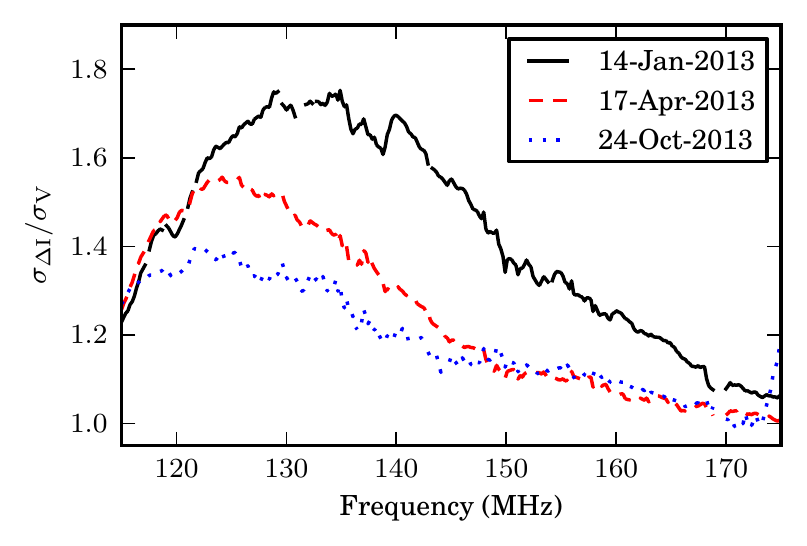}
\caption{The ratio of the RMS of differential Stokes I images ($\sigma_{\mathrm{\Delta I}}$) to those of Stokes V images ($\sigma_{\mathrm{V}}$), as a function of frequency for three observations. Consecutive sub-bands 195 kHz apart are used for the difference. The ratio is always greater than unity, implying there is an excess of noise in Stokes I as compared to the thermal noise dominated Stokes V. Sub-bands containing strong RFIs have been removed.}
\label{fig:excess}
\end{figure}

Another way to estimate the noise properties directly from the Stokes I parameter is to take the difference between two Stokes I images separated by a small frequency interval. All other signals from the sky, e.g. foregrounds and cosmological signal, should almost be the same in the two channels. The point spread function changes by only 0.1 percent over 0.2 MHz. Hence, the difference between two consecutive frequency channels should be dominated by the thermal noise, especially after the brightest discrete foreground sources have been subtracted. In principle, the noise properties obtained from the differential Stokes I images should be very close to those obtained from Stokes V. However, we find the Stokes I differential noise to be higher, as shown in Fig. \ref{fig:excess} where we plot the ratio of their RMS values for three different nights of observations. We call this additional noise in the Stokes I images the ``excess noise". The excess noise could originate from the following sources:
\begin{enumerate}
\item Convolution of residual sources with the chromatic PSF
\item Ionospheric scintillation
\item Calibration and foreground removal artefacts
\end{enumerate}
We perform several tests and simulations to study properties and causes of the excess noise. The potential sources, i.e. a chromatic PSF and ionospheric scintillation will be discussed in Section 4, whereas calibration artefacts will be discussed in Section 5.

The excess noise can not be removed by the foregrounds fitting algorithms which are used to remove faint sources and the diffuse foregrounds after subtracting the bright sources. Most of these algorithms separate the foregrounds from the 21 cm signal based on their smooth frequency spectra (\cite{Chapman2015} and references therein). The excess noise is uncorrelated even on small frequency separations of 0.2 MHz, and hence it can not be easily removed by standard foreground removal methods that expect spectrally smooth foregrounds.

\subsection{Suppression of the diffuse foregrounds}
The second systematic effect that we observe in the data is a suppression of the diffuse foregrounds, which occurs in the process of removal of discrete sources. Synchrotron and free-free emissions from our own galaxy constitute the diffuse foregrounds. These diffuse foregrounds are difficult to model and computationally expensive to include in the sky model for the direction dependent calibration in \texttt{SAGECal}. We remove them at a later stage based on their presumed smooth frequency spectra \citep{Harker2009, Chapman2012, Chapman2013}. Therefore, our sky model for \texttt{SAGECal} contains only discrete sources, whereas the observed data contains  also the diffuse foregrounds in total intensity as well as the linear polarization \citep{Jelic2014, Jelic2015}. A consequence of the difference between the true sky and the calibration sky model could be to suppress structures that are not part of the model, absorbing them in gains applied to the restricted calibration sky model and potentially lead to excess power elsewhere in the image or on different spatial or frequency scales.   

\begin{figure*}
\centering
\includegraphics{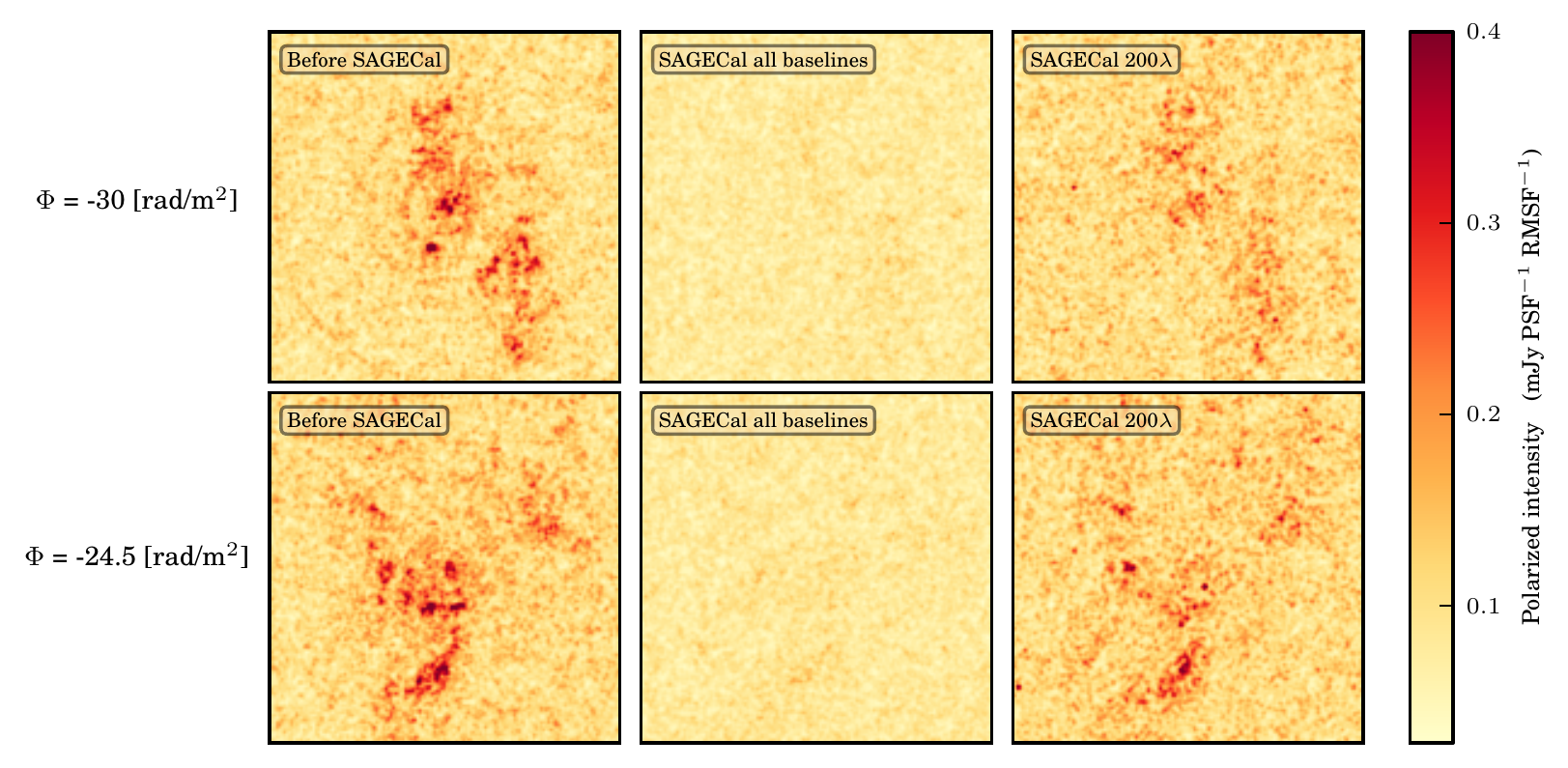}
\caption{Suppression of the diffuse foregrounds: uniform weighted, 4 degree polarized intensity maps for the following cases - i) before subtraction of discrete sources (first column), ii) after source subtraction using \texttt{SAGECal} (second column) and iii) after subtracting sources using baselines only longer than 200 wavelengths in calibration (third column). The top and bottom rows correspond to Faraday depths of -30 and -24.5 rad/m$^2$. The diffuse foregrounds are suppressed during the source subtraction because they are not included in the sky model. They can partially be recovered by excluding short baselines in calibration, but this results into an enhanced noise. The bright discerete sources present in the first column have been removed by \texttt{SAGECal} in other columns.}
\label{fig:suppression}
\end{figure*}

The suppression of the diffuse foregrounds is not easy to notice in Stokes I images because they are dominated by bright discrete sources and confusion noise. However, the suppression of the polarized diffused foregrounds can be easily seen, because not many discrete sources are polarized. The first two columns in Fig. \ref{fig:suppression} show the diffuse foregrounds in polarized intensity before and after the source subtraction, and the suppression in the latter case is self-evident. We show polarized intensity maps at two Faraday depths $(\Phi)$ of -30 and -24.5 obtained by rotation measure synthesis \citep{Brentjens2005}. The diffuse foregrounds appear on large angular scales where a detection of the 21 cm signal is also most promising \citep{Zaroubi2012, Chapman2013, Patil2014a}. Therefore, our concern is that a suppression in the diffuse foregrounds could mean a suppression of the 21 cm signal as well. A solution for mitigating the suppression of the diffuse foregrounds and the 21 cm signal is to exclude short baselines in the calibration. One can use only baselines longer than a certain baseline length and still obtain the gain solutions for all stations. Previously, \cite{Jelic2015} have used only baselines longer than 800 wavelengths in the calibration to minimize the suppression of the diffuse foregrounds. We use baselines longer than 200 wavelengths to obtain station gains but subtract the sky model sources on all baselines. As shown in the third column in Fig. \ref{fig:suppression}, this reduces the suppression of the diffuse foregrounds. One should note that the first and the third columns in Fig. \ref{fig:suppression} do not look exactly the same because the bright, largely instrumentally polarized, point sources present in the left panels have been subtracted using \texttt{SAGECal} in the right panels.

\section{Properties of the excess noise}
We performed several tests with an aim of investigating properties and ultimately the origin of the excess noise. Results of these tests are presented in this section.

\subsection{Angular power spectrum}
The angular power spectrum can be a useful tool in identifying causes of the excess noise. One should expect higher power on smaller angular scales if either sidelobes of sources due to the chromatic point source function (PSF) or ionospheric scintillation is the dominant cause of the excess noise. Sidelobes of unsubtracted sources are not perfectly subtracted in a sub-band difference due to the chromatic nature of the PSF \citep{Vedantham2012, Morales2012, Parsons2012}. The PSF is chromatic because the uv coordinate or the spatial frequency $u$ corresponding to a baseline scales with frequency $f$ as
\begin{equation} \label{eq:u}
u = \frac{bf}{c},
\end{equation}
where $b$ is the physical length of the baseline and $c$ is the speed of light. The rate of change of the uv coordinate with frequency, i.e.
\begin{equation} \label{eq:du}
\frac{du}{df} = \frac{b}{c},
\end{equation}
is larger at longer baselines. Therefore, we expect the power spectrum of the excess noise to increase with the baseline length, if a chromatic PSF were the dominant cause of the noise. Similarly, ionospheric scintillation noise shows more power on longer baselines \citep{Vedantham2015, Vedantham2016}. 

\begin{figure*}
\centering
\includegraphics[scale=1.25]{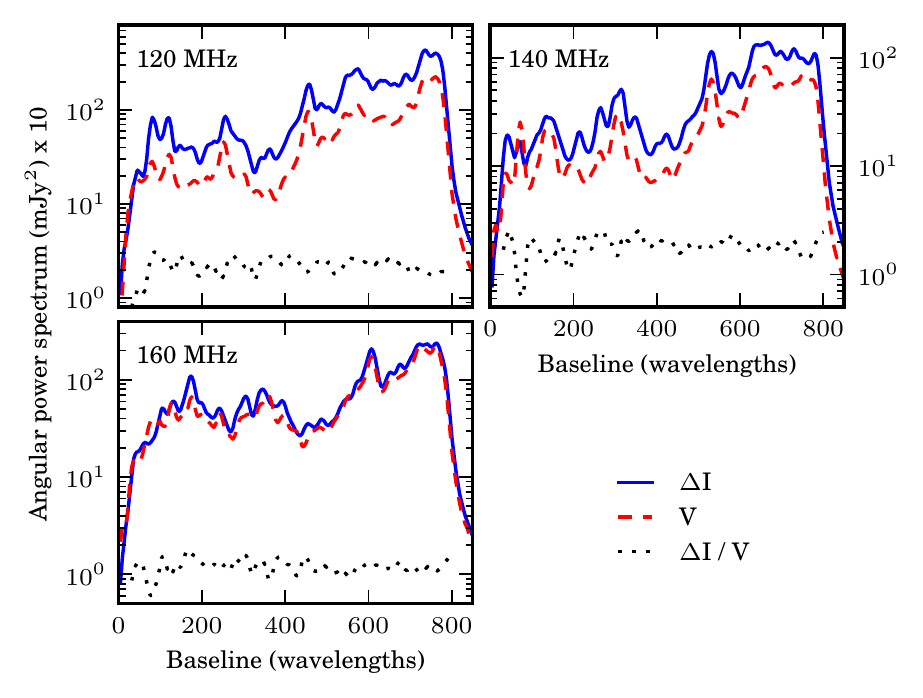}
\caption{Angular power spectrum of the excess and thermal noise for the observation on 17 April 2013. The ratio of the two remains constant irrespective of the baseline length. The power spectra have been multiplied by 10 for the convenience of plotting their ratio in the same plot.}
\label{fig:powspec}
\end{figure*}

We compute the azimuthally averaged angular power spectrum of the excess noise by Fourier transforming the differential Stokes I images and then squaring their magnitude. In Fig. \ref{fig:powspec}, we show the power spectrum of the excess noise as a function of baseline length for the observation on 17 April 2013. We also show the power spectrum of the thermal noise from Stokes V. The ratio of the power spectrum of the excess noise to that of the thermal noise remains  constant as a function of the baseline length. Therefore, we conclude that sidelobes of the unsubtracted sources and ionospheric scintillation are unlikely to be the dominant sources of the excess noise. This is in agreement with \cite{Vedantham2016} where it is shown that scintillation noise is confined to the wedge-like structure in the two dimensional power spectrum similar to smooth spectral foregrounds. 

\subsection{Contribution due to the chromatic PSF}
We estimate the contribution of sidelobes of discrete sources to the excess noise as follows.

The observed Stokes I signal in a frequency sub-band can be expressed as
\begin{equation} \label{eq:i1}
i_1 = s_1 * p_1 + n_{i1},
\end{equation}
where $s_1$ is the original signal from the sky, $p_1$ is the PSF, $n_{i1}$ is the thermal noise in Stokes I, and $*$ denotes a convolution operation. Taking a Fourier transform,
\begin{equation} \label{eq:I1}
I_1 = S_1 \times P_1 + N_{i1},
\end{equation}
where a capital letter denotes the Fourier transform of the respective quantity in equation (1). For Stokes V,
\begin{equation} \label{eq:V1}
V_1 = N_{v1},
\end{equation}
as we assume that the Stokes V contains only the thermal noise. Similarly, for a consecutive sub-band,
\begin{equation} \label{eq:I2}
I_2 = S_2 \times P_2 + N_{i2},
\end{equation}
\begin{equation} \label{eq:V2}
V_2 = N_{v2}.
\end{equation}
For a 195 kHz separation between two consecutive sub-bands, we assume that the signal from the sky does not change, i.e.
\begin{equation} \label{eq:S}
S = S_1 \approx S_2.
\end{equation}
The difference between the two sub-bands then becomes
\begin{equation} \label{eq:dI}
dI = I_1 - I_2 = S dP + N_{i1} - N_{i2},
\end{equation}
where $dP = P_1 - P_2$. We can compute the power spectrum of the differential Stokes I as
\begin{equation} \label{eq:dI2}
\left< \left| dI \right| ^2 \right> = \left| S \right|^2 \left| dP \right|^2 + \left< \left| N_{i1} \right|^2 \right> + \left< \left| N_{i2} \right|^2 \right>.
\end{equation}
Equation (10) follows from (9) because the thermal noise realizations at different sub-bands do not correlate. Similarly, for Stokes V,
\begin{equation} \label{dV2}
\left< \left| dV \right|^2 \right> = \left< \left| V_1 - V_2 \right|^2 \right> = \left< \left| N_{v1} \right|^2 \right> + \left< \left| N_{v2} \right|^2 \right>.
\end{equation}
The noise in Stokes I and V should be statistically identical, implying $\left< \left| N_{i1} \right|^2 \right> = \left< \left| N_{v1} \right|^2 \right>$ and $\left< \left| N_{i2} \right|^2 \right> = \left< \left| N_{v2} \right|^2 \right>$. Therefore, subtracting equation (11) from (10),
\begin{equation} \label{eq:dI_dV}
\left< \left| dI \right| ^2 - \left| dV \right| ^2 \right> = |S|^2 \left| dP \right| ^2,
\end{equation}
where the power spectrum of the signal from the sky $\left| S \right|^2$ is obtained using
\begin{equation} \label{S2}
\frac{\left| I_1 \right| ^2 - \left| V_1 \right| ^2}{\left| P_1 \right| ^2}  = \frac{\left| S \right|^2  \left| P_1 \right|^2 + \left| N_{i1} \right| ^2 - \left| N_{v1} \right| ^2}{\left| P_1 \right| ^2} = \left| S \right|^2.
\end{equation}

\begin{figure}
\centering
\includegraphics{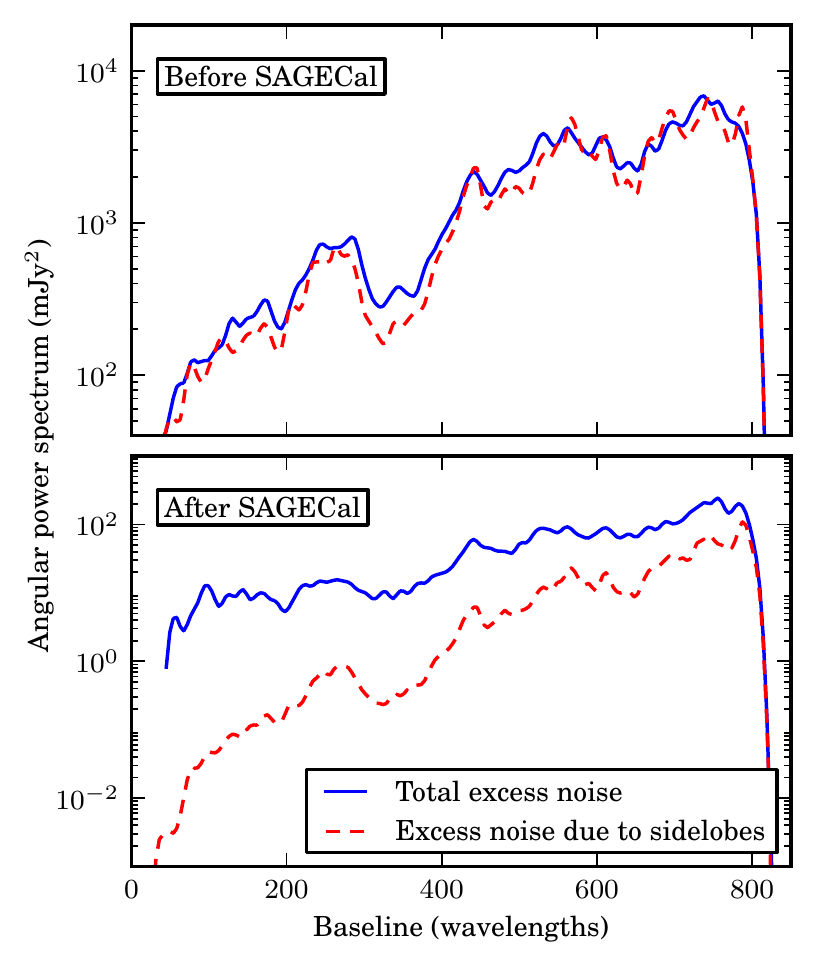}
\caption{Comparison of the total observed differential excess noise in differential Stokes I images with the differential sidelobe noise due to the chromatic PSF. Top panel: differential sidelobes account for the total excess noise before the direction dependent (DD) calibration and source subtraction with \texttt{SAGECal}. Bottom panel: The total excess noise is much higher than the contribution due to the differential sidelobes after the DD calibration.}
\label{fig:diffpsf}
\end{figure}

The left hand side of equation \ref{eq:dI_dV} is the power spectrum of the observed excess noise. Whereas the right hand side is the contribution of sidelobes of sources due to the chromatic PSF. Equation \ref{eq:dI_dV} implies that in an ideal case, where the sky signal does not change in consecutive sub-bands, nor other effects contribute such as the ionosphere or imperfect calibration,  the excess noise should be same as the differential sidelobe noise. We compute the power spectra of Stokes I, V and the PSF using uniform weighted images produced by  \texttt{ExCon}. The PSF images are produced by replacing all visibility data points by unity. We use the PSF at the center of the field in this test, assuming that the PSF does not vary significantly towards different directions. Fig. \ref{fig:diffpsf} shows the observed total excess noise and estimated contribution of the sidelobe noise, i.e. the right hand side of equation \ref{eq:dI_dV}, computed before and after the direction dependent calibration and source subtraction. The differential sidelobes amount to the total observed excess noise before source subtraction. However, it is only a small fraction of the excess noise after source subtraction. This suggests that the excess noise might have been introduced in the data during the source subtraction, and we will discuss this in detail in Section 5. 

\subsection{Correlation with the ionosperic scintillation}
In this subsection, we study any possible correlation of the excess noise with the ionospheric conditions. The ionosphere introduces stochastic phase fluctuations in the low frequency radio signals. \cite{Vedantham2015, Vedantham2016} have studied the scintillation noise due to ionospheric diffraction of discrete sources in the case of widefield interferometry. We expect the ionospheric scintillation noise to be higher when the diffractive scale is shorter \citep{Vedantham2015, Vedantham2016}. 

\begin{figure}
\centering
\includegraphics{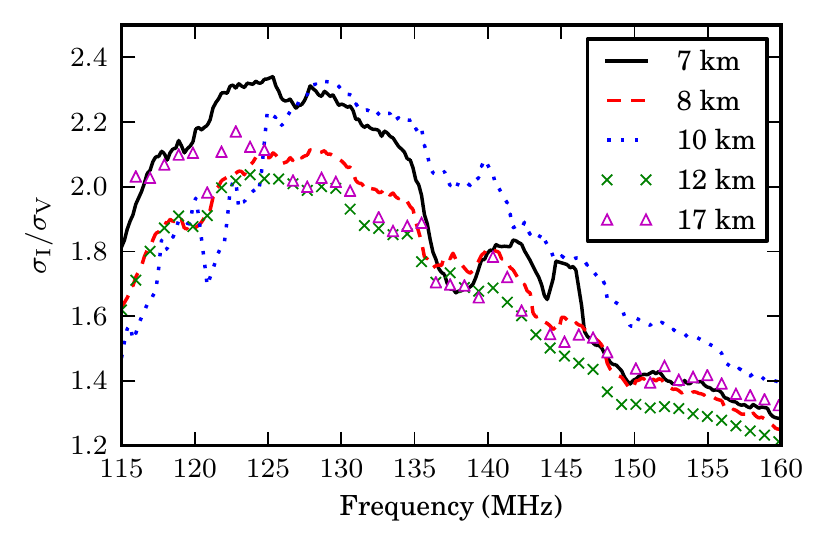}
\caption{The ratio of the RMS of \texttt{SAGECal} residuals in Stokes I to Stokes V as a function of frequency for different diffractive scales in the ionosphere observed on different nights. The diffractive scales are mentioned at 150 MHz. The shorter the diffractive scale, the higher the ionospheric scintillation noise. However, the noise in the data does not show an obvious anti-correlation with the diffrective scale.}
\label{fig:ionosphere}
\end{figure}

We briefly discuss here how we compute the diffractive scales from the data, but a more detailed description will follow in an upcoming paper (Mevius et al., in review). For each baseline, we compute the time series of the phase difference between the direction independent gain solutions of the pair of stations forming the baseline. We then compute the structure function which is the variance of the time series of the phase difference as a function of the baseline length. The structure function is fit to a power law, and it is expected to have a  power law index of $\frac{5}{3}$ for a Kolmogorov-type turbulence. The diffractive scale is the baseline length at which the phase variance is 1 radian$^2$. In Fig. \ref{fig:ionosphere}, we show the ratio of Stokes I to Stokes V RMS for different nights of observations with different diffractive scales. We do not find any obvious anti-correlation between the excess noise and the diffractive scale in the ionosphere. This again confirms our conclusion based on the angular power spectrum of the excess noise that the ionosphere is unlikely to be the dominant cause of the excess noise. 

We should note that we have seen an anti-correlation between the ionospheric diffractive scale and the noise before the direction dependent calibration and source subtraction in our other target field towards 3C196 which contains brighter sources (Mevius et al., in review). This effect might be difficult to see in the NCP field which does not contain bright sources. Furthermore, the traveling ionospheric disturbances are prominent on time scales of few minutes, and their effect is likely removed from the NCP data during the direction dependent calibration.

\section{Simulations}
In this section, we test whether the direction dependent calibration can introduce an excess noise using simulations of the calibration and source subtraction process where effects of the chromatic PSF and ionosphere can be eliminated. The simulated mock data-sets contain discrete sources, diffuse foregrounds and thermal noise. \texttt{SAGECal} is then used to obtain station gains and subtract the discrete sources. The steps involved in the simulations are as follows:

\begin{enumerate}
\item The brightest 25 sources are selected from the NCP sky model and their visibilities are predicted. The NCP sky model is constructed from the observed data and contains sources within a radius of 20 degrees around the NCP. The selected brightest 25 sources are located within a radius of 7 degrees from the NCP, and their flux densities range from 5 to 0.24 Jy.
\item  We predict the Stokes I visibilities of the simulated diffuse foregrounds from \cite{Jelic2008, Jelic2010}  multiplied with a time averaged primary beam of LOFAR. The RMS flux density of these diffuse foregrounds is normalized to 5 mJy/PSF, i.e. 7 K of brightness temperature. We do not know the brightness temperature of the diffuse foregrounds in the NCP field in total intensity, but we have assumed it to be ten times the brightness temperature of the observed polarized diffuse foregrounds in the field.
\item The thermal noise of RMS 1.5 Jy per visibility is simulated at the resolution of 10 s, 183 kHz at 135 MHz. This results into a RMS noise of 0.83 mJy per sub-band image for a 13 hours long observation, which is comparable to the observed noise in Stokes V images in the data.
\item Visibilities of the discrete sources, diffuse foregrounds and the thermal noise are added to form a mock dataset.
\item \texttt{SAGECal} is used to calibrate the station gains and remove discrete sources from the simulated data. We cluster the simulated 25 sources in 21 directions for which the station gain solutions are obtained. We keep the number of directions small so that the calibration remains an over-determined system\footnote{The 64 LOFAR stations in the Netherlands form 2016 baselines. Therefore, one can solve for gains in maximum 31 directions for 64 stations in a snapshot. We use 5 to 20 minutes time intervals in \texttt{SAGECal} which provide more constrains.}.
\end{enumerate}

While predicting visibilities for discrete sources, we increase their fluxes by 5 percent. This is equivalent to station gains being higher than their expected values. This way, we ensure that the actual values of gain solutions in the calibration are not the same as the initial values used in calibration iterations. Such absolute scaling of fluxes does not affect the end result. However, if we were to vary relative fluxes of sources grouped within a cluster, that would affect the common solution for that group of sources. In the following subsections, we present the results of different tests performed with the simulations.

\subsection{Different noise realizations of one sub-band}
\begin{figure}
\centering
\includegraphics{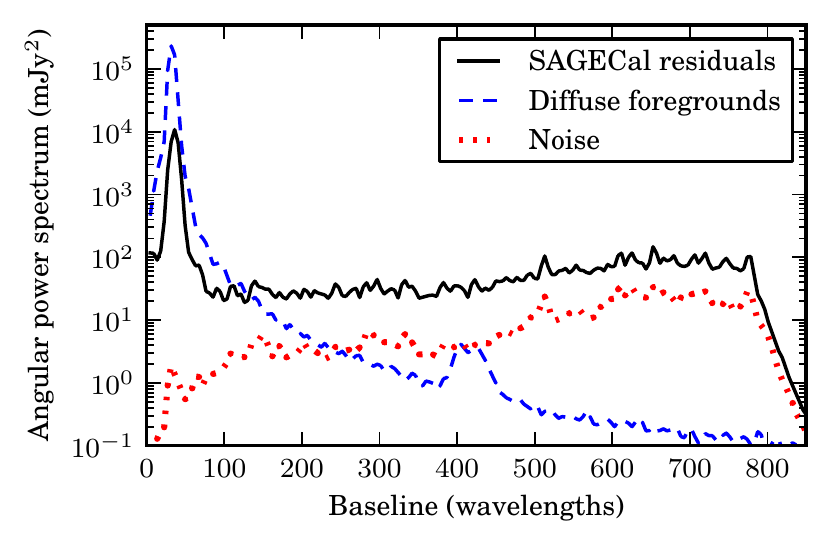}
\includegraphics{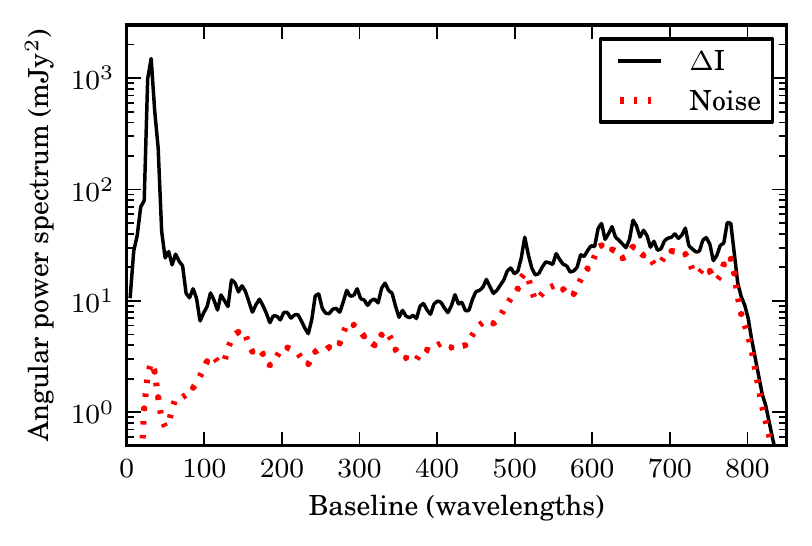}
\caption{Results from multiple noise realizations of one frequency sub-band. Top panel: Angular power spectra of the input diffuse foregrounds, thermal noise and \texttt{SAGECal} residuals after source subtraction. The diffuse foregrounds are suppressed at short baselines in residuals, whereas long baselines show  excess power above the thermal noise. Bottom panel: differential residuals ($\Delta I$) between different noise realizations, which are higher than the thermal noise. The simulated data contains 25 discrete sources (5-0.24 Jy), the diffuse foregrounds (7 K) and the thermal noise (0.83 mJy/PSF).}
\label{fig:manynoise}
\end{figure}

Here, we simulate multiple realizations of the mock data for one frequency sub-band at 135 MHz. Different realizations contain the same discrete and diffuse foregrounds but different realizations of the thermal noise. The advantage of this test is that we exclude effects of the chromatic PSF in this analysis. Ideally, we expect the discrete sources to get perfectly subtracted and the diffuse foregrounds with the thermal noise to be left as residuals. However, as shown in the top panel of Fig. \ref{fig:manynoise}, we find an excess of power in the residuals at baselines longer than 200 wavelengths, i.e. the discrete sources are not perfectly subtracted. Additionally, the power at short baselines is suppressed, i.e. the diffuse foregrounds are partially removed during the source subtraction. As the diffuse foregrounds remain the same in different data realizations, we expect the difference between the residuals of different realizations to be consistent with the thermal noise. However, as shown in the bottom panel of Fig. \ref{fig:manynoise}, we see an excess of flux in the differential residuals of different realizations. The power spectrum of the differential residuals resembles thermal noise only at baselines longer than 100 wavelengths. At shorter baselines, the diffuse foregrounds affect the power spectrum of the residuals.

\begin{figure}
\centering
\includegraphics{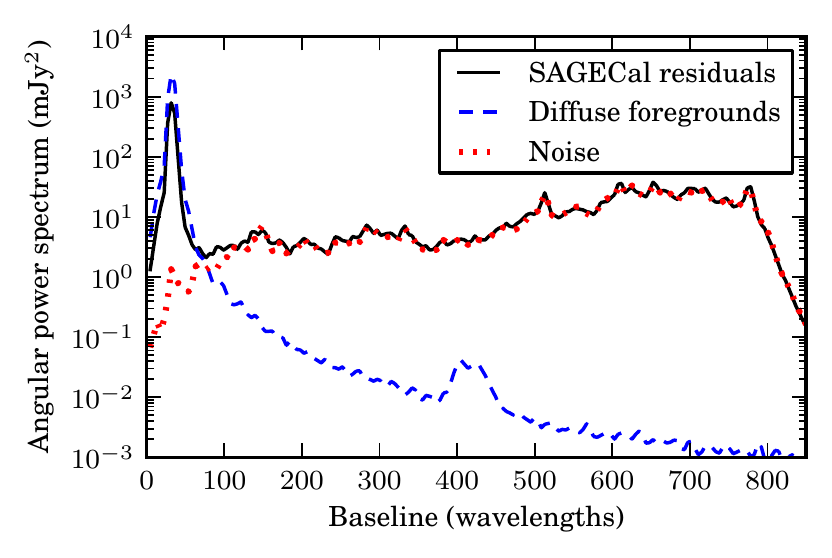}
\caption{Simulation results, same as Fig. \ref{fig:manynoise}, except here the brightness of the diffuse foregrounds is reduced by 10 times. The foreground suppression is reduced, and the excess noise has disappeared as compared to Fig. \ref{fig:manynoise}, showing that these systematic effects are functions of the unmodeled flux due to the diffuse foregrounds.}
\label{fig:lowfg}
\end{figure}

We find that both the suppression of the diffuse foregrounds and the excess noise depend on the brightness of the diffuse foregrounds which are not part of the sky model. In Fig. \ref{fig:lowfg}, we show the results when the intensity of the diffuse foregrounds is reduced by ten times to have a RMS of 0.7 K. The suppression of foregrounds is reduced, and the residuals reach the thermal noise at long baselines. This test shows that both the foreground suppression and the excess noise problems occur when the sky model used in self-calibration and source subtraction is incomplete. Additionally, the intensity of these problems depends on the missing flux in the model. \citep{Barry2016} suggested unmodelled foregrounds convolved with a chromatic PSF as the source of variations in calibration solutions and an excess noise. However, as evident from this test, unmodelled flux in itself could be sufficient to cause variations in calibration solutions. 

\subsection{Multiple \texttt{SAGECal} runs on the same realization of simulation}
\begin{figure}
\centering
\includegraphics{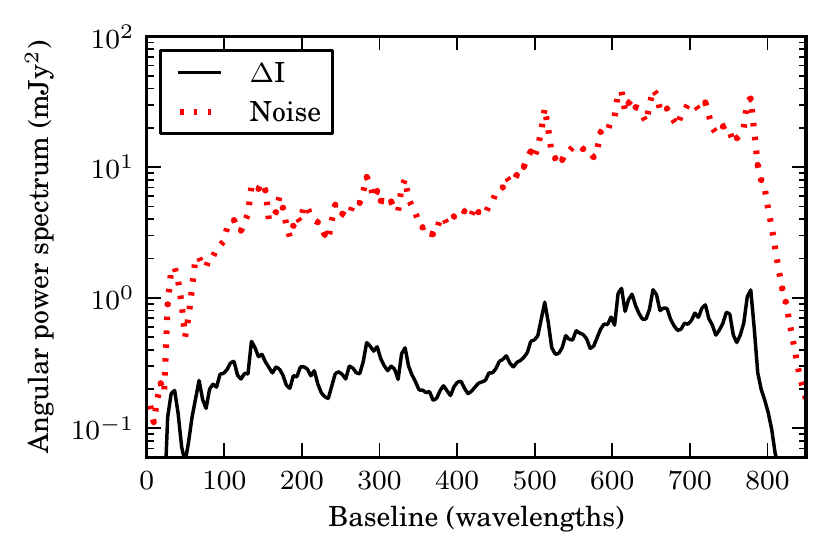}
\caption{Results from multiple \texttt{SAGECal} runs on one realization of the simulation. The difference between residuals of different runs ($\Delta I$) is 10 percent of the thermal noise, and it has the same power spectrum as the thermal noise.}
\label{fig:samenoise}
\end{figure}

In this subsection, we study how any randomization used in the calibration may result in an excess noise. In every expectation maximization step in \texttt{SAGECal}, the order in which the station gains in different directions are solved, is randomized to reduce the systematic errors in the solver. However, the final solution in every run of \texttt{SAGECal} is expected to reach the global minimum in the likelihood space. Therefore, the difference between the residuals of different calibration runs on the same data should be near zero. However, we observe differential residuals when one realization of the simulated data is run multiple times through the calibration and source removal. For a simulation containing discrete sources in the flux range 5 to 0.24 Jy and diffuse foregrounds of RMS brightness temperature 0.7 K, the differential noise is 10 percent of the thermal noise. However, the level of this excess noise depends on the relative fluxes of the discrete sources and the diffuse foregrounds as summarized in Table 2. As shown in Fig. \ref{fig:samenoise}, the power spectrum of the differential noise resembles that of the thermal noise just as observed in the real data, unless the unmodelled flux dominates on certain baselines which was the case in Fig. \ref{fig:manynoise}. 

We believe that the calibration procedure finds different solutions in different runs on the same data and can not actually reach the global minimum in the likelihood space because of the unmodeled flux due to the diffuse foregrounds. This hypothesis could in principle be verified by sampling the likelihood space of calibration parameters. However, this is computationally very expensive for our parameter space of high dimensions (21 directions x 64 stations x 2 polarization components).

\begin{table}
\centering
\caption{The differential noise ($\Delta I$) in residuals of multiple \texttt{SAGECal} runs on the same realization of the simulated data for different levels of discrete and diffuse foregrounds. The diffuse foregrounds are mentioned in flux densities of RMS/PSF and in RMS brightness temperature in parentheses. The differential noise in residuals is mentioned as a percentage of the thermal noise.}
\begin{tabular}{l c r}
\hline
Discrete sources & Diffuse foregrounds & $\Delta I$ / Noise \\
\hline
5 to 0.24 Jy & 5 mJy (7 K) & 130\% \\
5 to 0.24 Jy & 0.5 mJy (0.7 K) & 10\% \\
0.5 to 0.24 Jy & 0.5 mJy (0.7 K) & 25\% \\
\end{tabular}
\label{table:randomization}
\end{table}

\section{Possible solutions to the foreground suppression and excess noise}
The simulations presented in Section 5 have shown a clear evidence that the excess noise and suppression of the diffuse foregrounds occur because of an incomplete model in self-calibration. We now discuss two possible solutions to mitigate these systematic errors.

\subsection{Excluding short baselines from calibration}
\begin{figure}
\centering
\includegraphics{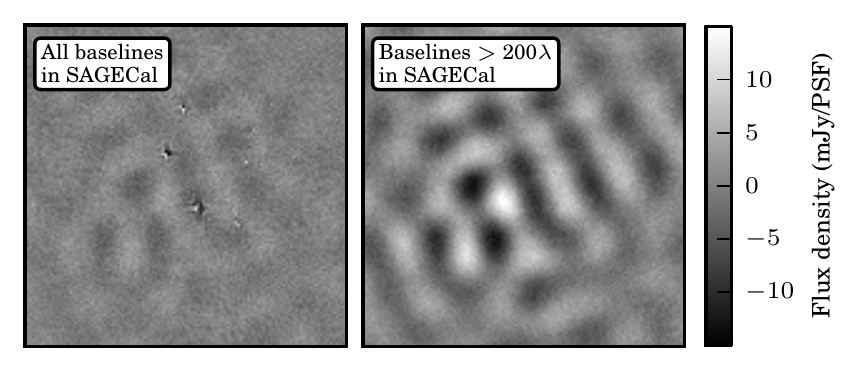}
\caption{Comparison of \texttt{SAGECal} residuals (uniform weighted, 10 degree images) when all baselines are used for calibration (left panel) and only baselines longer than 200 wavelengths are used (right panel). When all baselines are used, the sky model is incomplete due to the missing diffuse foregrounds. As a result, the diffuse foregrounds are suppressed and the discrete sources are imperfectly subtracted. Excluding short baselines in the calibration resolves both of these issues.}
\label{fig:fgimg}
\end{figure}

\begin{figure}
\centering
\includegraphics{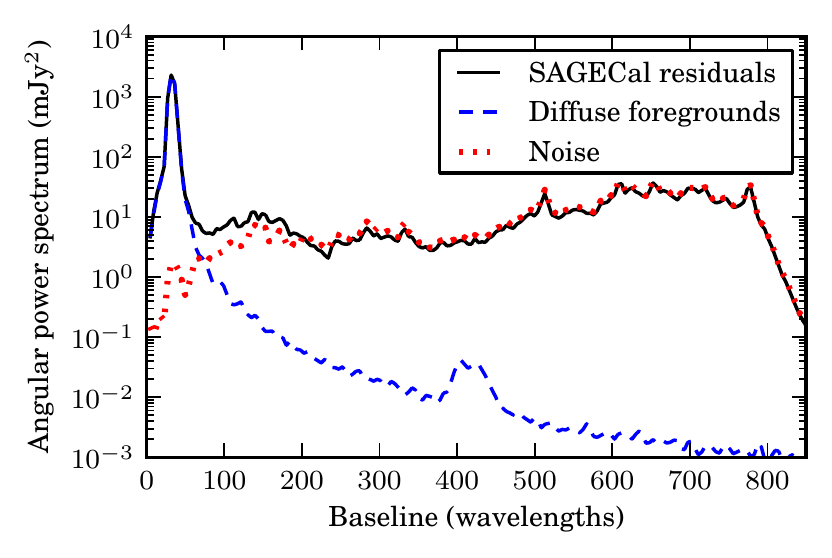}
\includegraphics{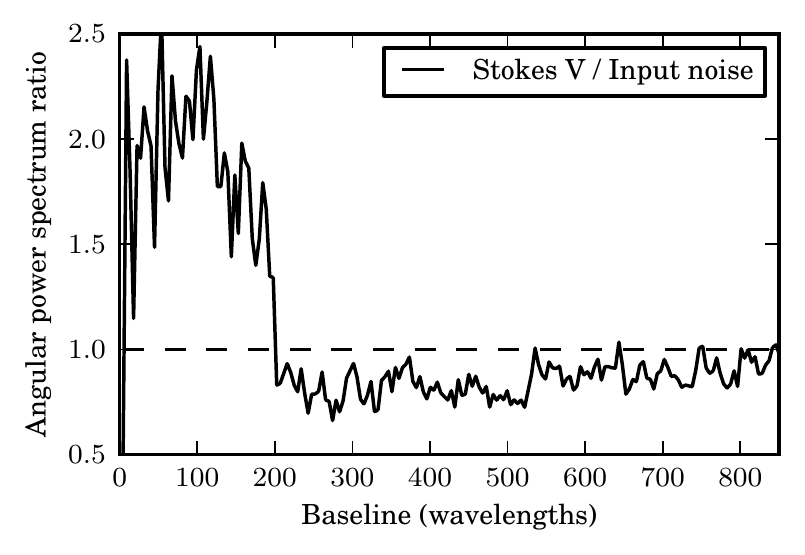}
\caption{Excluding baselines shorter than 200 wavelengths in calibration. Top panel: the \texttt{SAGECal} residuals contain the diffuse foregrounds without any suppression. Additionally, the residuals reach the thermal noise at longer baselines implying perfect removal of the discrete sources and no excess noise. Bottom panel: the ratio of the power spectrum of the thermal noise after source removal (Stokes V) to that of the input noise. The noise is enhanced by a factor of 2 on baselines which were excluded from calibration.}
\label{fig:leverage}
\end{figure}

The diffuse foregrounds are not part of the sky model, but they are dominant only on short baselines. Their brightness is negligible at baselines longer than 200 wavelengths as compared to the discrete sources in total intensity in the NCP field. We can use baselines only longer than 200 wavelengths to obtain gain solutions for all stations and then subtract sources on all baselines. In such a case, the diffuse foregrounds would affect the self-calibration at a much reduced level. In Fig. \ref{fig:fgimg}, we compare \texttt{SAGECal} residuals when all and only long baselines are used for the calibration of the 25 brightest sources in the NCP field in the presence of 7 K diffuse foregrounds. In the former case, the suppression of the diffuse foregrounds and residuals of the discrete sources is evident. Both of these issues are mitigated in the latter case. The top panel of Fig. \ref{fig:leverage} shows the same phenomenon in the form of angular power spectra. When the short baselines are excluded in the calibration, the diffuse foregrounds remain untouched in the residuals at short baselines. Additionally, there is no excess noise at long baselines because the discrete sources are perfectly removed. Excluding short baselines, however, has a severe disadvantage as it enhances the noise on the excluded baselines. In the bottom panel of Fig. \ref{fig:leverage}, we plot the ratio of the power spectrum of the noise (Stokes V) after source subtraction to that of the input noise. The noise on the excluded baselines is boosted by a factor of 2 in power. A mathematical derivation of this phenomenon is given in the appendix for interested readers. The enhancement of noise implies a loss in sensitivity on short baselines, i.e. large angular scales, which otherwise would have been most promising for a detection of the 21 cm signal \citep{Patil2014a, Zaroubi2012}. As evident in the lower panel of Fig. \ref{fig:leverage}, the thermal noise is suppressed by 10 percent on long baselines which are used for the calibration. However, this suppression would affect any further analysis because these long baselines will only be used for the calibration but not for a detection of the 21 cm signal.

\subsection{Simultaneous multi-frequency calibration}
The calibration is often performed on one frequency sub-band at a time due to computing and memory constraints. This gives the station-gain solutions a partial freedom to vary independently at different sub-bands, producing an excess noise which is uncorrelated along frequency, as also shown by \cite{Barry2016}. As seen in our data as well as simulations, the power spectrum of this excess noise is similar to that of the thermal noise. \cite{Trott2016b} also reached to the same conclusion in the context of bandpass calibration. 

The primary beam as well as any ionspheric effects vary smoothly with frequency. Therefore, a parametric calibration can be obtained for a large bandwidth instead of independent gain solutions at each sub-band. \citep{Barry2016} suggested fitting a low-order polynomial to gain solutions along frequency or averaging calibration solutions of multiple interferometric elements. Alternatively, \cite{Yatawatta2015a} have proposed a regularization which enforces smoothness on the calibration solutions to a degree depending on the chosen value of the regularization parameter. As a result, the errors on the station gains are reduced, although the theoretical limit based on the thermal noise can not be reached due to the model incompleteness. We also believe that a simultaneous multi-frequency calibration should reduce the suppression in the diffuse foregrounds. The unmodeled flux due to the diffuse foregrounds changes significantly from 115 to 170 MHz. Therefore, the suppression should be reduced if the entire or a significant fraction of the bandwidth is simultaneously used to constrain the calibration solutions. We leave a more detailed analysis of the multi-frequency calibration for future work.

\section{Conclusions}
The LOFAR Epoch of Reionization (EoR) project aims to detect the redshifted 21 cm emission from neutral hydrogen from redshift 6 to 11. It is crucial to control the systematic errors for a signal detection, because the foregrounds are several orders of magnitude brighter than the expected signal. In this paper, we have studied two systematic biases observed in the residual LOFAR EoR data after calibration and subtraction of bright discrete foreground sources: i) a suppression in the diffuse emission and ii) excess of noise beyond the thermal component. These biases occur because of the direction dependent calibration with an incomplete sky model, and they are potential obstacles in a signal detection for the following reasons. 
\begin{enumerate}
\item Both the diffuse foregrounds and the 21 cm signal are easiest to detect on large angular scales, and the suppression of the former might imply a suppression of the 21 cm signal as well.
\item The excess noise implies a loss in sensitivity and an additional bias in a measurement of the power spectrum of the 21 cm signal. Furthermore, the excess noise would not be removed by the foreground removal methods which remove spectrally smooth signals.
\end{enumerate}
The differential noise between two closely spaced frequency bins after removing the bright sources from the data, is higher than the thermal noise. We call this additional noise: ``excess noise". We have performed tests to study properties of the excess noise and identify its causes. The angular power spectrum of the excess noise resembles that of the thermal noise, i.e. it shows the same power on all baselines. The chromatic point spread function (PSF) and ionospheric scintillation would have shown increasing power with the baseline length. We have estimated that the contribution of sidelobes of the unsubtracted sources due to the chromatic PSF is only a small fraction of the excess noise. The excess noise in different observations does not show any obvious correlations with the diffractive scales in the ionosphere on respective nights. Therefore, we establish that the chromatic PSF and ionosphere scintillation can not be the dominant causes of the excess noise.

We use simulated data-sets to study the systematic errors that could be produced by the calibration and source subtraction algorithms. Just like the real data, the discrete sources are removed by modeling them, calibrating the  LOFAR station gains in their directions and then subtracting the sources. The calibration minimizes the difference between the data and the model by adjusting the station gains. In this process, the diffuse foregrounds are suppressed, because they are not part of the model. This also results in imperfect removal of the discrete sources. The source residuals are partially uncorrelated in multiple noise realizations of the simulated data. This could explain the excess noise in the difference between two frequency bins in the actual data which contain uncorrelated realizations of the thermal noise. The angular power spectrum of the excess noise resembles that of the thermal noise in the simulations, just as it does in the actual data, and its magnitude depends on the amount of flux that is included in the sky model relative to the amount of flux that is excluded in the model.

We discuss two possible solutions to the observed systematic biases. Firstly, short baselines where the diffuse foregrounds are dominant, can be excluded from the calibration. This ensures that the diffuse foregrounds and the 21 cm signal are not suppressed. However, it enhances the noise on the excluded baselines, implying a poor sensitivity on large angular scales where a detection of the 21 cm signal otherwise would have been most promising. Secondly, we believe a better solution would be to use multi-frequency constraints to enforce spectral smoothness on the calibration parameters. Our future efforts are going to  be focused on that front \citep{Yatawatta2015a}. 

\section{Acknowldegement}
AHP and SZ would like to thank the Lady Davis Foundation and The Netherlands Organization for Scientific Research (NWO) VICI grant for the financial support. LVEK and BKG acknowledge the financial support from the European Research Council under ERC-Starting Grant FIRSTLIGHT - 258942. AGdB, SY, MM and VNP acknowledge support by the ERC for project 339743 (LOFARCORE). VJ acknowledges the NWO for the financial support under VENI grant - 639.041.336. ITI was supported by the Science and Technology Facilities Council [grant number ST/L000652/1]. The Low-Frequency  Array (LOFAR) was designed and constructed by ASTRON, the Netherlands Institute for Radio Astronomy, and has facilities in several countries, which are owned by various parties (each with their own funding sources) and are collectively operated by the International LOFAR Telescope (ILT) foundation under a joint scientific policy.

\bibliographystyle{mn2e}
\bibliography{eor.bib}

\appendix
\section{Leverage as a diagnostic in calibration}
In this appendix, we provide a mathematical proof of the enhancement of noise on baselines which are excluded in calibration. We use Leverage, a well known concept in regression analysis, to study the performance of calibration. Leverage \citep{Cook1982} can be loosely described as the change in the predicted value based on the data model used, due to the change in the data used for estimating the calibration parameters. In nonlinear regression, Jacobian Leverage \citep{Laurent1992,Laurent1993} is widely used  \citep{Neugebauer1996}. Here we apply it to study calibration. We adopt a case deletion model in regression \citep{Ross1987} to study the situation where only a subset of baselines (or data points) are used for calibration \citep{Yatawatta2015b}.

\subsection{Radio interferometric calibration}
Here, we give a brief overview of the data model used in radio interferometric calibration \citep{Hamaker1996, Thompson2007}. In interferometry, the correlated signal from $p$-th and $q$-th stations, ${\bf {\sf V}}_{pq}$  is given by
\begin{equation} \label{vispq}
{\bf {\sf V}}_{pq}= \sum_{i=1}^{K}  {\bf {\sf J}}_{pi} {\bf {\sf C}}_{pqi} {\bf {\sf J}}_{qi}^{H} + {\bf {\sf N}}_{pq},
\end{equation}
where ${\bf {\sf J}}_{pi}$ and ${\bf {\sf J}}_{qi}$ are the Jones matrices describing errors along the direction of source $i$ at stations $p$ and $q$, respectively. The matrices represent the effects of the propagation medium, the beam shape and the receiver. There are $K$ sources in the sky model and the noise matrix is given as ${\bf {\sf N}}_{pq}$. The contribution from the $i^{\mathrm{th}}$ source on baseline $pq$ is given by the coherency matrix $C_{pqi}$. We estimate the Jones matrices ${\bf {\sf J}}_{pi}$ for $p\in[1,R]$ and $i\in[1,K]$, during calibration and calculate the residuals by subtracting the predicted model (multiplied with the estimated Jones matrices) from the data. The vectorized form of (\ref{vispq}), ${\bmath v}_{pq}=\mathrm{vec}({\bf {\sf V}}_{pq})$  can be written as 
\begin{equation} \label{vecvispq}
{\bmath v}_{pq}= \sum_{i=1}^K {\bf {\sf J}}_{qi}^{\star}\otimes {\bf {\sf J}}_{pi} \mathrm{vec}({\bf {\sf C}}_{pqi}) + {\bmath n}_{pq}
\end{equation}
where ${\bmath n}_{pq}=\mathrm{vec}({\bf {\sf N}}_{pq})$. Depending on the time and frequency interval within which calibration solutions are obtained, we can stack up all cross correlations within that interval as
\begin{equation}
{\bmath d}=[\mathrm{real}({\bmath v}^T_{12})\ \mathrm{imag}({\bmath v}^T_{12})\ \mathrm{real}({\bmath v}^T_{13})\ldots \ldots \mathrm{imag}({\bmath v}^T_{(R-1)R})]^T
\end{equation}
where  ${\bmath d}$ is a vector of size $N\times 1$ of real data points. Thereafter, we have the data model 
\begin{equation} \label{obs}
{\bmath d}=\sum_{i=1}^{K}{\bmath s}_i({\bmath \theta}) + {\bmath n}
\end{equation}
where ${\bmath \theta}$ is the real parameter vector (size $M\times 1$) that is estimated by calibration. The contribution of the $i$-th known source on all data points is given by ${\bmath s}_i({\bmath \theta})$ (size $N\times 1$). The noise vector is given by  ${\bmath n}$ (size $N\times 1$). The parameters ${\bmath \theta}$ are the elements of ${\bf {\sf J}}_{pi}$-s, with real and imaginary parts considered separately. 

The maximum likelihood (ML) estimate of ${\bmath \theta}$ under zero mean, white Gaussian noise is obtained by minimizing the least squares cost 
\begin{equation} \label{mltheta}
\widehat{\bmath \theta}=\argmin_{\bmath \theta} \|{\bf d}- \sum_{i=1}^{K}{\bf s}_i({\bmath \theta})\|^2
\end{equation}
as done in current calibration approaches \citep{Boonstra2003, vanderVeen2005, Kazemi2011} and this is improved by using a weighted least squares estimator to account for errors in the sky model \citep{Kazemi2013a}. The Cramer-Rao Lower Bound is used to find a lower bound to the variance of $\widehat{\bmath \theta}$ \citep{Zmuidzinas2003, vanderTol2007, Wijnholds2009, Kazemi2012}. However, relating this lower bound to the residual ${\bmath d}-\sum_{i=1}^{K}{\bmath s}_i(\widehat{\bmath \theta})$ is not simple. Instead, we propose Leverage to quantify errors on the residuals.

\subsection{Leverage}
Consider a nonlinear regression model
\begin{equation}
{\bmath y}={\bmath m}({\bmath \theta}) + {\bmath n},
\end{equation}
where $\bmath y$ is a $N \times 1$ data vector, $\bmath n$ is the $N \times 1$ noise vector and $\bmath m(\bmath \theta)$ is a nonlinear function of the $M \times 1$ parameter vector $\bmath \theta$. The residual vector $\bmath r(\bmath \theta)$ is given by
\begin{equation}
\bmath r(\bmath \theta) = \bmath y - \bmath m(\bmath \theta).
\end{equation}
The estimated value of $\bmath \theta$ using (weighted) least squares is given by $\widehat{\bmath \theta}$ and the predicted value based on the estimated parameters is given by $\widehat{\bmath y}= \bmath m(\widehat{\bmath \theta})$. Now consider perturbing the data by $b \bmath f$ where $\bmath f$ ($N \times 1$) is any arbitrary vector and $b$ is a real scalar. Let us call the perturbed data as $\bmath y_b$ and the estimated value of $\bmath \theta$ using the perturbed data as $\widehat{\bmath \theta}_b$. The predicted value using $\widehat{\bmath \theta}_b$ is denoted by $\widehat{\bmath y}_b$. We define the leverage vector as \citep{Laurent1992}
\begin{equation}
{\bmath g} \buildrel \triangle \over = \lim_{b \to 0} \frac{1}{b} \left(\widehat{\bmath y}_b - \widehat{\bmath y} \right),
\end{equation}
and for (weighted) least squares estimation, we define Jacobain leverage as \citep{Laurent1993}
\begin{eqnarray}\label{Jlev}
{\bmath \Gamma}({\bmath \theta}) &\buildrel \triangle \over =& {\bmath \eta}_{\bmath \theta} \left( {\bmath \eta}_{\bmath \theta}^T {\bmath \eta}_{\bmath \theta}  - \sum_{i=1}^{N} {\bmath r}^i \left( ({\bmath \eta}^i)_{{\bmath \theta}{\bmath \theta}} \right) \right)^{-1} {\bmath \eta}_{\bmath \theta}^T, \nonumber\\
{\bmath \eta}_{\bmath \theta} &=& \frac{\partial}{\partial {\bmath \theta}^T} {\bmath m}({\bmath \theta}), \nonumber\\
({\bmath \eta}^i)_{{\bmath \theta} {\bmath \theta}} &=& \frac{\partial^2}{\partial {\bmath \theta}\partial {\bmath \theta}^T} {\bmath m}^i({\bmath \theta}),
\end{eqnarray}
where ${\bmath r}^i$ is the $i$-th element in ${\bmath r}({\bmath \theta})$ and ${\bmath m}^i({\bmath \theta})$ is the $i$-th element in ${\bmath m}({\bmath \theta})$. We see that ${\bmath \eta}_{\bmath \theta}$ is a matrix of size $N\times M$ and $({\bmath \eta}^i)_{{\bmath \theta} {\bmath \theta}}$ is a matrix of size $M\times M$. Once we have ${\bmath \Gamma}({\bmath \theta})$ ($N\times N$) matrix, and also the estimated parameters $\widehat{\bmath \theta}$, given any arbitrary vector ${\bf f}$ ($N\times 1$), we can find ${\bmath g}={\bmath \Gamma}(\widehat{\bmath \theta}) {\bmath f}$ \citep{Laurent1992}.

Now consider the case when the model is a summation of $L$ nonlinear functions, and that each function depends only on a subset of parameters (also called as partially separable), i.e.,
\begin{equation}\label{sepmod}
{\bmath m}({\bmath \theta}) = \sum_{i=1}^{L} {\bmath h}_i({\bmath \theta}_i)
\end{equation}
with ${\bmath \theta} = [{\bmath \theta}_1^T,{\bmath \theta}_2^T,\ldots]^T$. Also assume that we are only interested in finding the diagonal values of ${\bmath \Gamma}({\bmath \theta})$. In this case, applying (\ref{Jlev}) to (\ref{sepmod}) yields 
\begin{equation}\label{grad}
{\bmath \eta}_{\bmath \theta} =
\left[ \begin{array}{cccc}
{\bmath \eta}_1 & {\bmath \eta}_1 & \ldots & {\bmath \eta}_L\\
\end{array} \right]
\end{equation}
where ${\bmath \eta}_i = \frac{\partial}{\partial {\bmath \theta}_i^T} {\bmath h}_i({\bmath \theta}_i)$ and
\begin{equation}\label{hess}
({\bmath \eta}_i^j)_{{\bmath \theta} {\bmath \theta}}=
\left[ \begin{array}{cccc}
{\bf {\sf H}}_1^j & {\bf {\sf 0}} & \ldots & {\bf {\sf 0}}\\
{\bf {\sf 0}} & {\bf {\sf H}}_2^j &  \ldots & {\bf {\sf 0}}\\
\vdots & \vdots &  \vdots & \vdots\\
{\bf {\sf 0}} & {\bf {\sf 0}} & \ldots & {\bf {\sf H}}_L^j \\
\end{array} \right]
\end{equation}
where
\begin{equation}
{\bf {\sf H}}_i^j = \frac{\partial^2}{\partial {\bmath \theta}_i\partial {\bmath \theta}_i^T} {\bmath h}_i^j({\bmath \theta}_i)
\end{equation}
with ${\bmath h}_i^j({\bmath \theta}_i)$ being the $j$-th element of ${\bmath h}_i({\bmath \theta}_i)$. Substituting (\ref{grad}) and (\ref{hess}) to (\ref{Jlev}) and only considering the block diagonal entries, we get
\begin{equation}\label{Jlevsum}
{\bmath \Gamma}({\bmath \theta})=\sum_{j=1}^L {\bmath \eta}_j \left( {\bmath \eta}_j^T {\bmath \eta}_j - \sum_{i=1}^{N} {\bmath r}^i {\bf {\sf H}}_j^i \right)^{-1} {\bmath \eta}_i^T 
\end{equation}
which can be used to get the diagonal entries of ${\bmath \Gamma}({\bmath \theta})$.

\subsection{Calibration with excluded data}
We consider the general case where a subset of data (baselines) are excluded during calibration. Consider ${\mathcal J}$ to be the set of indices of excluded data points in (\ref{obs}). Assume the total ignored data points to be $R$, $0\le R < N$. Following \cite{Ross1987}, we modify (\ref{obs}) as
\begin{equation} \label{mD}
{\bmath d}=\sum_{i=1}^{K}{\bmath s}_i({\bmath \theta}) + {\bf {\sf D}} {\bmath \gamma} + {\bmath n},
\end{equation}
where ${\bf {\sf D}}$ ($N\times R$) is a matrix whose $i^{\mathrm{th}}$ column has $1$ at the ${\mathcal J}^i$-th location and the rest of the entries in the column are $0$. We introduce an additional parameter vector ${\bmath \gamma}$ ($R\times 1$) into the data model. Normally  ${\bmath \gamma}$ is called the cross-validatory residual. The effect of these slack variables is to nullify the constraints introduced by the data points indexed by  the set ${\mathcal J}$. If ${\bmath \theta}_r=[{\bmath \theta}^T {\bmath \gamma}^T]^T$ are the augmented parameters ($M+R$), calibration gives us
\begin{equation} \label{mlD}
\widehat{\bmath \theta}_r = \argmin_{{\bmath \theta},{\bmath \gamma}} \|{\bf d}- \sum_{i=1}^{K}{\bf s}_i({\bmath \theta}) - {\bf D} {\bmath \gamma}\|^2
\end{equation}
even though we do not explicitly solve for ${\bmath \gamma}$. Therefore, the calibration with excluded data (\ref{mlD}) estimates $M+R$ parameters using $N$ constraints, while calibration with all data (\ref{mltheta})  estimates $M$ parameters using $N$ constraints. In both cases, the useful set of parameters is still ${\bmath \theta}$ of size $M$.

Now we apply (\ref{Jlevsum}) for the data model in (\ref{mD}), where we have $L=K+2$, with $K$ nonlinear functions ${\bmath s}_j({\bmath \theta}_j)$ (parameters ${\bmath \theta}_j$), one linear function ${\bf {\sf D}} {\bmath \gamma}$ (parameters ${\bmath \gamma}$) and noise ${\bmath n}$. 
\begin{enumerate}
\item ${\bmath s}_j({\bmath \theta}_j)$: The values for ${\bmath \eta}_j$ and ${\bf {\sf H}}_j^i$ for each $j$ can be calculated using (\ref{vecvispq}), and since this is quadratic, both ${\bmath \eta}_j$ and ${\bf {\sf H}}_j^i$ are non zero, but they are sparse.
\item ${\bf {\sf D}} {\bmath \gamma}$: Since this is linear in ${\bmath \gamma}$, ${\bmath \eta}_j={\bf D}$ and ${\bf {\sf H}}_j^i={\bf {\sf 0}}$.
\item ${\bmath n}$: For noise, we do not have any parametrization, and therefore, we assume both ${\bmath \eta}_j$ and ${\bf {\sf H}}_j^i$ to be matrices with random entries. We notice the following for the computation of the leverage:
\end{enumerate}
Considering the aforementioned three cases, we see that (i) and (iii) are always present, regardless of calibration using the full dataset $R=0$ or a subset of baselines ($R>0$). In other words, (i) and (iii) contribute to (\ref{Jlevsum}) in both cases. Moreover, the contribution (iii) is not dependent on ${\bmath \theta}$ and therefore is uniform if the noise ${\bmath n}$ is uniformly distributed. The interesting case is (ii), when $R>0$. The contribution to (\ref{Jlevsum}) can be written as
\begin{equation}
{\bmath \Gamma}_d = {\bf {\sf D}}\left({\bf {\sf D}}^T{\bf {\sf D}}-{\bf {\sf 0}}\right)^{-1}{\bf {\sf D}}^T = {\bf {\sf D}} {\bf {\sf D}}^T = \widetilde{\bf {\sf I}}
\end{equation}
where $\widetilde{\bf {\sf I}}$ is a diagonal matrix with $1$-s at the locations  given by ${\mathcal J}$ and the rest of the entries $0$. To sum up: if the $i^{\mathrm{th}}$ diagonal entry of ${\bmath \Gamma}({\bmath \theta})$ calculated with the estimate $\widehat{\bmath \theta}$ using the full dataset is $\Gamma^{ii}(\widehat{\bmath \theta})$, then this value changes to $\Gamma^{ii}(\widehat{\bmath \theta})+1$ for the case where the $i^{\mathrm{th}}$ datapoint is excluded during calibration. The excluded baselines have an increase in leverage by $1$. Therefore, the error in the residuals is enhanced on the excluded baselines. The only way to minimize this error is to minimize the variance of estimated parameters, $\widehat{\bmath \theta}$, or in other words, find the global minimum point in the parameter space.

\end{document}